\documentclass[twocolumn]{webofc}

\usepackage[varg]{txfonts}   
\newcommand{\sib}[1]{{\sc Sibyll}~#1\xspace}
\newcommand{\qgsii}{{\sc Qgsjet~II-04}\xspace}
\newcommand{\eposlhc}{{\sc Epos-lhc}\xspace}

\newcommand{\Xmax}{$X_{\rm max}$\xspace}
\newcommand{\erange}[2]{\ensuremath{10^{#1}-10^{#2}~{\rm eV}}\xspace}
\newcommand{\lgE}[1]{\ensuremath{10^{#1}~{\rm eV}}\xspace}
\usepackage{subfig}
\usepackage{mathrsfs}

\usepackage{tikz}
\usetikzlibrary{positioning,shapes,shadows,arrows,backgrounds,curvilinear,intersections,fit, matrix}
\usetikzlibrary{plotmarks,patterns}
\usepackage{pgfplots}

\usepgflibrary{plotmarks}
\pgfplotsset{compat=1.5}
\usepgfplotslibrary{external} 
\usepackage{hyperref}

\begin{document}
\title{Testing Model Predictions of Depth of Air-Shower Maximum and Signals in Surface Detectors using Hybrid Data of the Pierre Auger Observatory}
%
%

\author{\firstname{Jakub} \lastname{Vícha}\inst{1}\fnsep\thanks{\email{vicha@fzu.cz}}
       \firstname{} \lastname{for Pierre Auger Collaboration}\inst{2}\fnsep\thanks{\email{spokespersons@auger.org}
}}

\institute{Institute of Physics of the Czech Academy of Sciences 
\and
           Observatorio Pierre Auger, Av. San Martín Norte 304, 5613 Malargüe, Argentina \\
           for full-author list see \href{https://www.auger.org/archive/authors_2022_10.html}{https://www.auger.org/archive/authors\_2022\_10.html}
          }

\abstract{%
  We present a method for testing the predictions of hadronic interaction models and improving their consistency with observed two-dimensional distributions of the depth of shower maximum, $X_\text{max}$, and signal at the ground level as a function of zenith angle. The method relies on the assumption that the mass composition is the same at all zenith angles, while the atmospheric shower development and attenuation depend on composition in a correlated way. In the present work, for each of the three leading LHC-tuned hadronic interaction models, we allow a global shift $\Delta X_\text{max}$ of the predicted shower maximum, which is the same for every mass and energy, and a rescaling $R_\text{Had}$ of the hadronic component at the ground level which is constant with the zenith angle.

We apply the analysis to 2297 events reconstructed with both the fluorescence and surface detectors of the Pierre Auger Observatory with energies $10^{18.5-19.0}$ eV and zenith angles below 60$^\circ$. Given the modeling assumptions made in this analysis, the best fit reaches its optimum value when shifting the $X_\text{max}$ predictions of hadronic interaction models to deeper values and increasing the hadronic signal. This change in the predicted $X_\text{max}$ scale alleviates the previously identified model deficit in the hadronic signal (commonly called the muon puzzle) but does not fully remove it. Because of the size of the adjustments $\Delta X_\text{max}$ and $R_\text{Had}$ and the large number of events in the sample, the statistical significance of need for these adjustments is large, greater than 5$\sigma_\text{stat}$, even for the combination of the systematic experimental shifts within 1$\sigma_\text{sys}$ that are the most favorable for the models.
}
\maketitle

\section{Introduction}
\label{intro}
The cosmic rays of ultra-high energy (above $10^{18}$~eV) are still of unknown origin and uncertain mass composition despite the huge effort of the community for more than 50 years since their discovery.
The main complications are their very low flux and uncertainties coming from the extrapolation of hadronic interactions that take place in the development of air-showers initiated by the primary cosmic rays.
The Pierre Auger Observatory \cite{PACosmicObservatory} provides an unprecedented large amount of high-quality air showers reconstructed using two detection techniques - Surface Detector (SD) and Fluorescence Detector (FD), providing an excellent opportunity to test predictions of hadronic interaction models.

\begin{figure}[h!]
\includegraphics[width=0.48\textwidth]{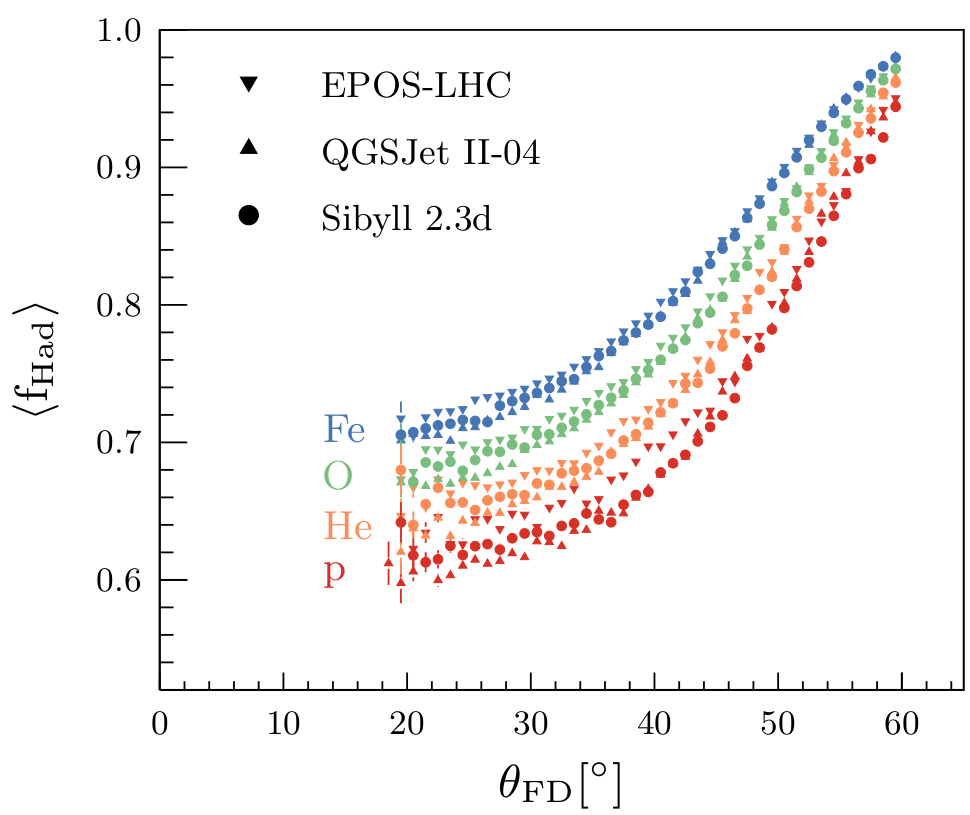}
\caption{Average fraction of the hadronic signal at 1000~m, $f_\text{Had}=S_\text{Had}/S(1000)$, as a function of the zenith angle reconstructed by the FD. Signals from four primaries (colors) of energies \erange{18.5}{19.0} and three models of hadronic interactions (markers) are shown.}
\label{fig:HadronicFraction}
\end{figure}

\begin{figure*}[h!]
\subfloat{\includegraphics[width=0.475\textwidth]{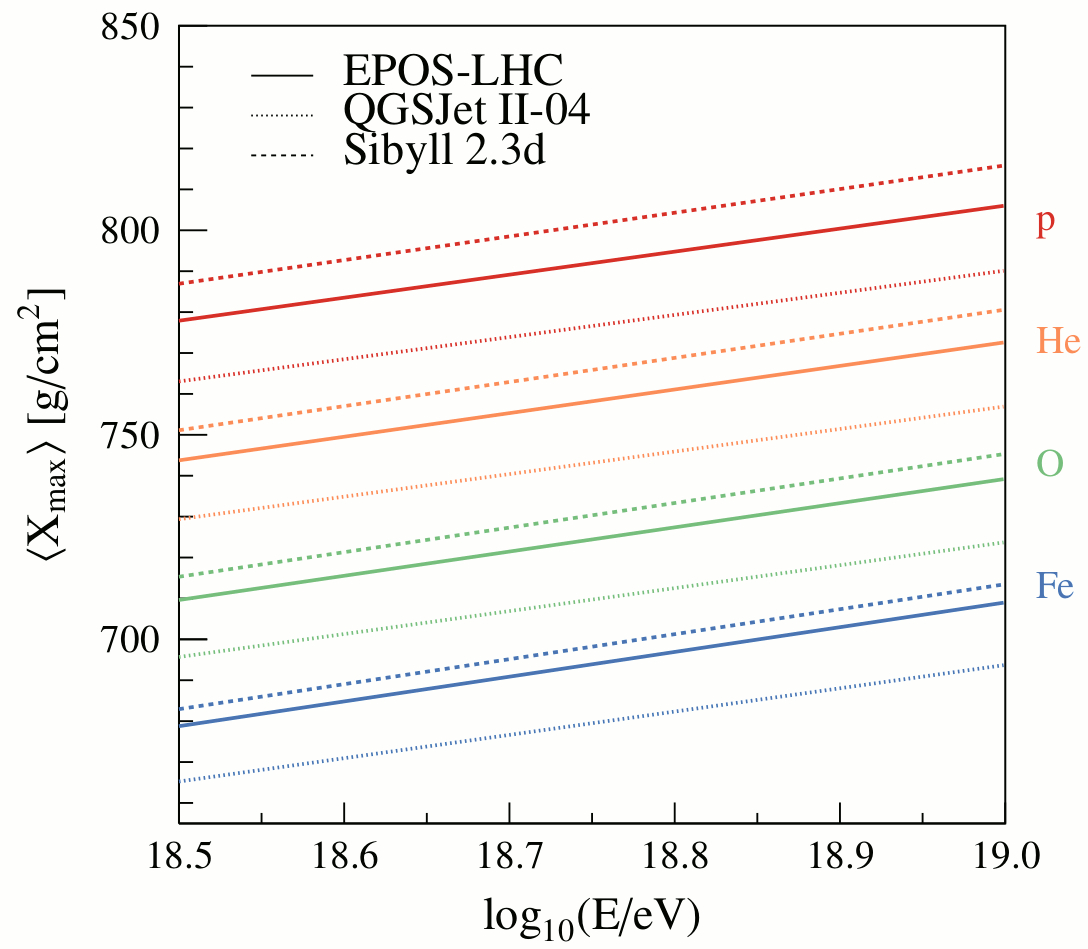}}
\hspace{0.5cm}
\subfloat{\includegraphics[width=0.5\textwidth]{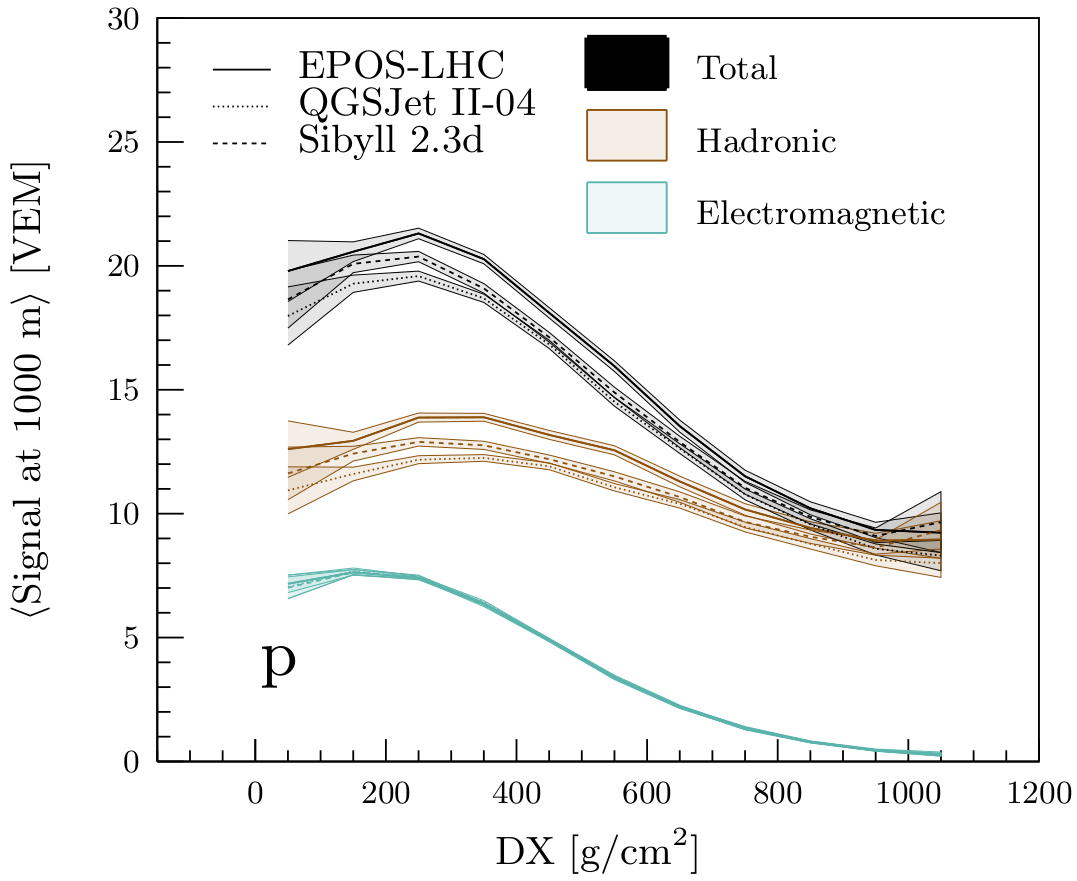}}
\caption{Left: Energy evolution of the mean of $X_{\rm max}$ distribution predicted for three hadronic interaction models and four primary species. The model \sib{2.3d} predicts on average by $\approx$7 g/cm$^{2}$ and $\approx$22 g/cm$^{2}$ deeper $\langle$\Xmax$\rangle$ than the models \eposlhc and \qgsii, respectively. Right: Decomposition of the total ground signal (black) into the hadronic (braun) and EM (turquoise) parts as predicted to depend on the distance of \Xmax to the ground ($DX$). The ground signal at 1000~m from the shower core is simulated for responses at the Pierre Auger Observatory to showers generated using three hadronic interaction models in case of protons of energies \erange{18.5}{19.0}.}
    \label{ModelDifferences}
\end{figure*}

The SD measurement of the ground signal estimates the shower size using the signal expected at 1000~m from the shower core, $S(1000)$ \cite{SDEnergySpectrum2020}.
In this work, the signal $S(1000)$ is assumed to be composed of the hadronic $S_{\rm Had}$ and electromagnetic $S_{\rm em}$ components.
The signal $S_{\rm Had}$ is induced by muons, and electromagnetic (EM) particles from muon decays and low-energy neutral pions according to the 4-component shower universality description \cite{ShowerUniversality},\cite{ShowerUniversality2}.
The signal $S_{\rm em}$ is induced by EM particles originating from high-energy neutral pions.
The muon component attenuates less with the zenith angle ($\theta$) than the EM component and its size is a measure of the primary mass.
As a consequence, the fraction of the hadronic signal to $S(1000)$ is increasing with $\theta$ and mass of the primary (protons p, and He, O, Fe nuclei), see Fig.~\ref{fig:HadronicFraction}.

The FD measurement of the deposited energy on the shower depth provides a precise estimation of the shower energy ($E_\text{FD}$) and depth of the shower maximum (\Xmax) \cite{Auger-LongXmaxPaper}.
The latter quantity is another measure of the mass of primary particle initiating the shower.

In this work, we use the showers reconstructed in both SD and FD to analyse simultaneously the mass composition of primary particles and deficiencies of Monte Carlo (MC) predictions on $S(1000)$ and \Xmax.
We test in this way predictions of three models of hadronic interactions \eposlhc \cite{EposLHC}, \qgsii\cite{Qgsjet}, \sib{2.3d} \cite{Sibyll} in the energy range around the ankle region \cite{SDEnergySpectrum2020}.

\subsection{Data Selection and Observables}
We select 2297 events detected with FD energies \erange{18.5}{19.0} and $\theta$ within 60$^\circ$ during the period 1. 1. 2004 - 31. 12. 2018.
These events survive high-quality selection criteria used to derive the SD energy spectrum \cite{SDEnergySpectrum2020} and FD analysis of \Xmax \cite{Auger-LongXmaxPaper,XmaxICRC19}.
These events are divided into 5 zenith-angle ranges containing nearly the same and sufficiently large number of events.

\begin{figure*}[h!]
\includegraphics[width=\textwidth]{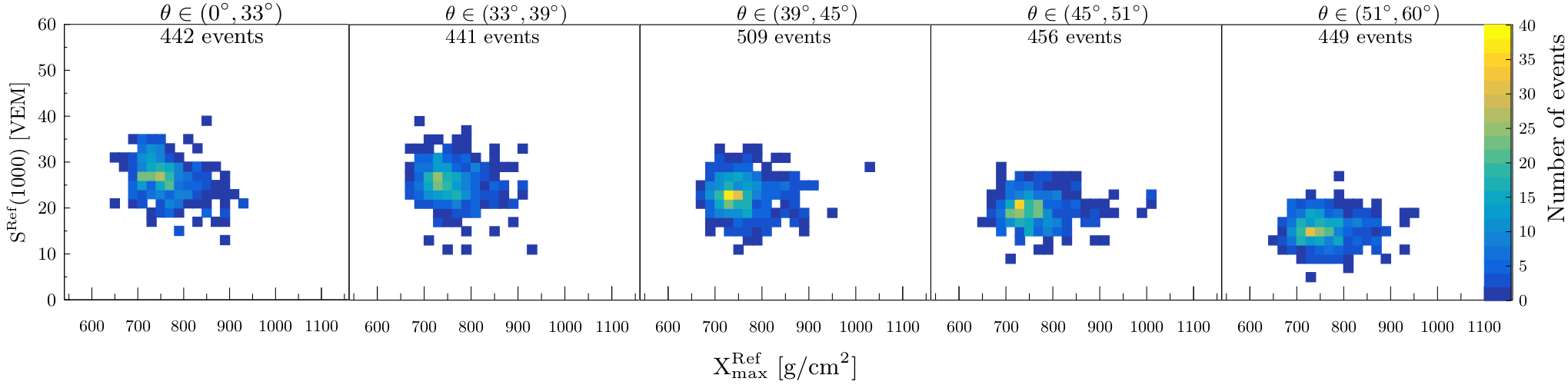}
\caption{The two-dimensional distributions of $S(1000)$ and \Xmax for data measured by the Pierre Auger Observatory in the energy range \erange{18.5}{19.0} and divided into five zenith-angle ($\theta$) bins.}
\label{fig:Data}
\end{figure*}

    \renewcommand{\arraystretch}{1.5}
  \begin{table*}
    \begin{center}
      \caption{Minimal values of the Likelihood-ratio expression from Eq.(\ref{LogLfunction}) for the best fits of three hadronic interaction models allowing different adjustments to the MC templates.}
      \begin{tabular}{|c||c|c|c|}
        \hline 
        $\mathscr{L}_\text{min}$ & \eposlhc & \qgsii & \sib{2.3d} \\ 
        \hline \hline 
        none & 2097.410 & 4650.070 & 2581.530 \\
        $\Delta X_{\rm max}$ & 764.090 & 1746.945 & 1060.795 \\
        $R_\text{Had}$ & 518.146 & 720.086 & 550.954 \\
        $R_\text{Had}$ \& $\Delta X_{\rm max}$ & 479.456 & 515.055 & 480.611 \\
        \hline
      \end{tabular}
      \label{TabLikelihoods}
    \end{center}
  \end{table*}
  
The observables $X_\text{max}^\text{Ref}$ and $S^\text{Ref}(1000)$ contain corrections for the energy evolution of \Xmax and $S(1000)$ using the FD energy as
  \begin{equation}
    S^{\rm Ref}(1000)=S(1000)\cdot \left( \frac{E^{\rm Ref}}{E_{\rm
        FD}} \right )^{1/B},
    \label{GroundSignalEq}
  \end{equation}
  and
  \begin{equation}
    X_{\rm max}^{\rm Ref}=X_{\rm max}+D\cdot {\rm log}_{10} \left(
    \frac{E^{\rm Ref}}{E_{\rm FD}} \right),
    \label{XmaxEq}
  \end{equation}
  where $B$ = 1.031 is the SD energy calibration parameter \cite{SDEnergySpectrum2020} and the elongation rate of a single primary $D = 58$~g/cm$^{2}$ is taken as the average value over the four primary particles and the three hadronic interaction models. The reference energy is set to $10^{18.7}$~eV.
  
\section{Adjustments to Monte Carlo Predictions}
\label{sec-1}
There are persistent inconsistencies in the mass interpretation of \Xmax measurements and ground-signal measurements \cite{TestingHadronicInteractions}, \cite{AmigaMuons}.
All these interpretations are based on the assumption of \Xmax scale, $\langle$ \Xmax $\rangle$ at given energy, predicted by a given model of hadronic interactions and driving this way an assumption on the mass composition.
Another assumption in these analyses is a tension between the data and simulations coming only from the lack of hadronic (muon) signal in the MC predictions.

Recently, a novel method considering insufficiencies not only in the predictions of ground signal, but also in the predicted \Xmax scale was applied to the data of the Pierre Auger Observatory \cite{MethodAugerDataICRC21}. 
The motivation for such a generalized approach is indicated on the left panel of Fig.~\ref{ModelDifferences} where the differences in model predictions of $\langle X_\text{max}\rangle$ are approximately energy and primary-mass independent.
On the right panel of Fig.~\ref{ModelDifferences}, the difference in the ground signal predicted by different models of hadronic interactions is stemming from the different scales of hadronic signal.
The EM part of signal is very universal within the applied shower-universality approach to the ground signal. 

\subsection{Fitting Method in Nutshell}

In this work, we show the application of this novel method to the data of the Pierre Auger Observatory considering simultaneously freedom in the mass composition and the predicted hadronic and \Xmax scale.
The freedom in the predicted hadronic signal is assumed to be a constant function of the zenith angle on the contrary to the more general assumptions made in \cite{MethodAugerDataICRC21}, where more details can be found. 

In short, we perform a composition fit simultaneously to five two-dimensional (2D) distributions of $X_\text{max}^\text{Ref}$ and $S^\text{Ref}(1000)$ corresponding to five zenith-angle ranges between 0$^\circ$ and 60$^\circ$, see Fig.~\ref{fig:Data}, with a combination of MC templates for four primary species (p, He, O, Fe).
A constant freedom in the predicted \Xmax scale (\Xmax=\Xmax+$\Delta X_\text{max}$) and hadronic scale ($S_\text{Had}=S_\text{Had}\cdot R_\text{Had}$) in these templates is considered.
The effect of the change of $X_{\rm max}$ on the ground signal is incorporated through the separate effects on the hadronic and EM signals estimated from the parameterized evolution of the mean ground signal parts with the distance of $X_{\rm max}$ to the ground ($DX=880{~\rm g/cm}^{2}/\cos(\theta)-X_{\rm max}$).
In this way, the total ground signal in a given $\theta$-range is estimated to be modified by about 7\% at most in case of a change of $X_{\rm max}$ by 50 g/cm$^{2}$.

The likelihood-ratio expression that is minimized in the method for a given model of hadronic interactions is of the form
  \begin{eqnarray}
    \forall n_{jz} > 0:&~~\mathscr{L}=\sum\limits_{z}\sum\limits_{j}\left(
    C_{jz}-n_{jz}+n_{jz}\cdot \ln\frac{n_{jz}}{C_{jz}} \right), \nonumber\\
    ~~~~\forall n_{jz} = 0:&~~\mathscr{L}=\sum\limits_{z}\sum\limits_{j}C_{jz},
    \label{LogLfunction}
  \end{eqnarray} 
  with the sums running over the 2D-bins $j$ and five $\theta$-bins $z$.  
  The number of showers measured in a 2D-bin $j$ and a $\theta$-bin $z$ is denoted by $n_{jz}$ and the predicted number of MC showers for the same bin by $C_{jz}$.  
These MC predictions at the reconstruction level are obtained from parameterization of MC templates for each of the three models of hadronic interactions.

\begin{figure*}
\includegraphics[width=\textwidth]{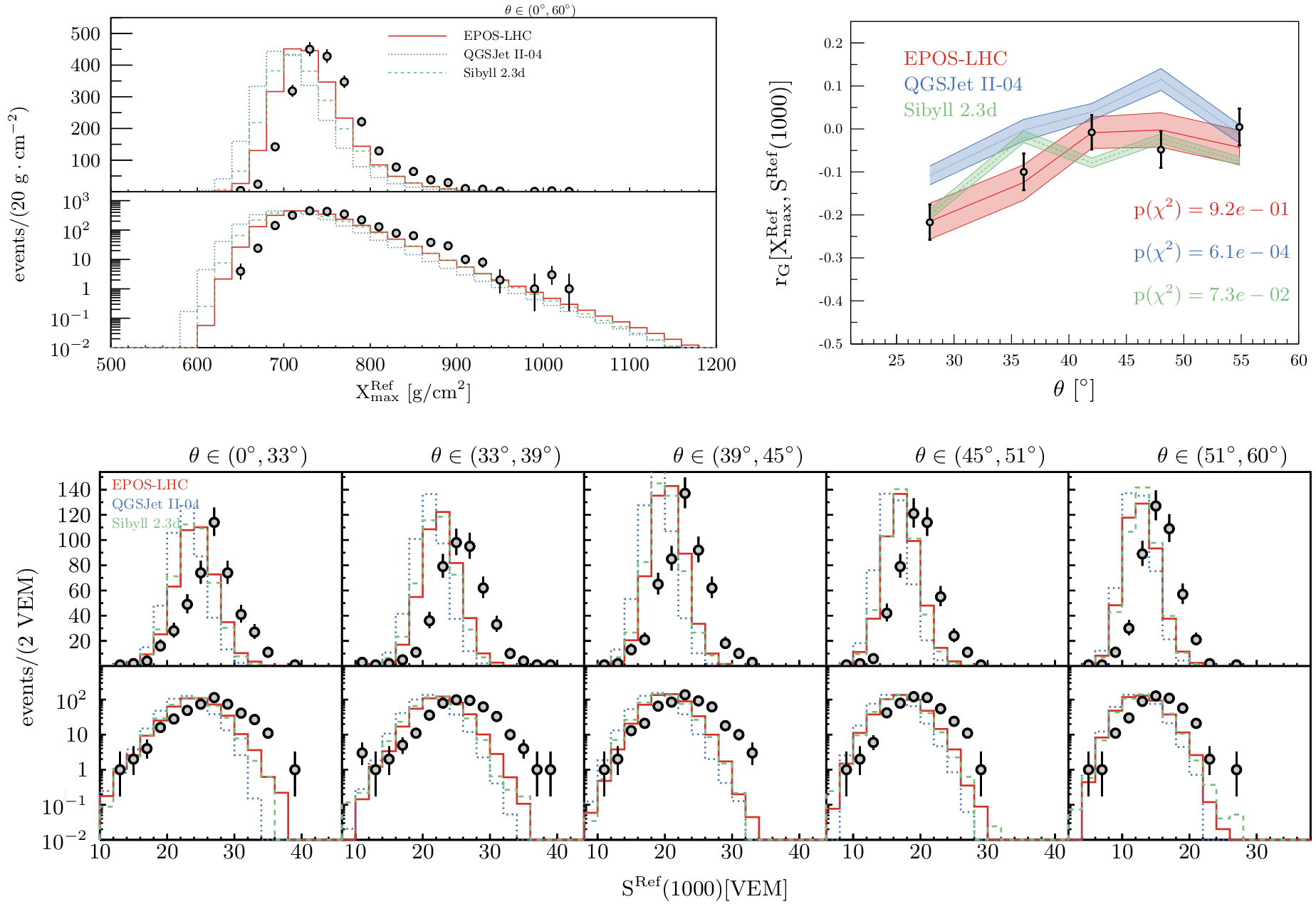}
\caption{Description of data (black) by the composition fits without any modification of the MC templates. The projected \Xmax distribution (top left), projected $S(1000)$ distributions for all five zenith-angle ($\theta$) ranges and the zenith-angle dependence of the correlation between \Xmax and $S(1000)$ are shown for the three models of hadronic interactions (lines).}
\label{fig:DataDescription_NoMods}
\end{figure*}

\begin{figure*}
\includegraphics[width=\textwidth]{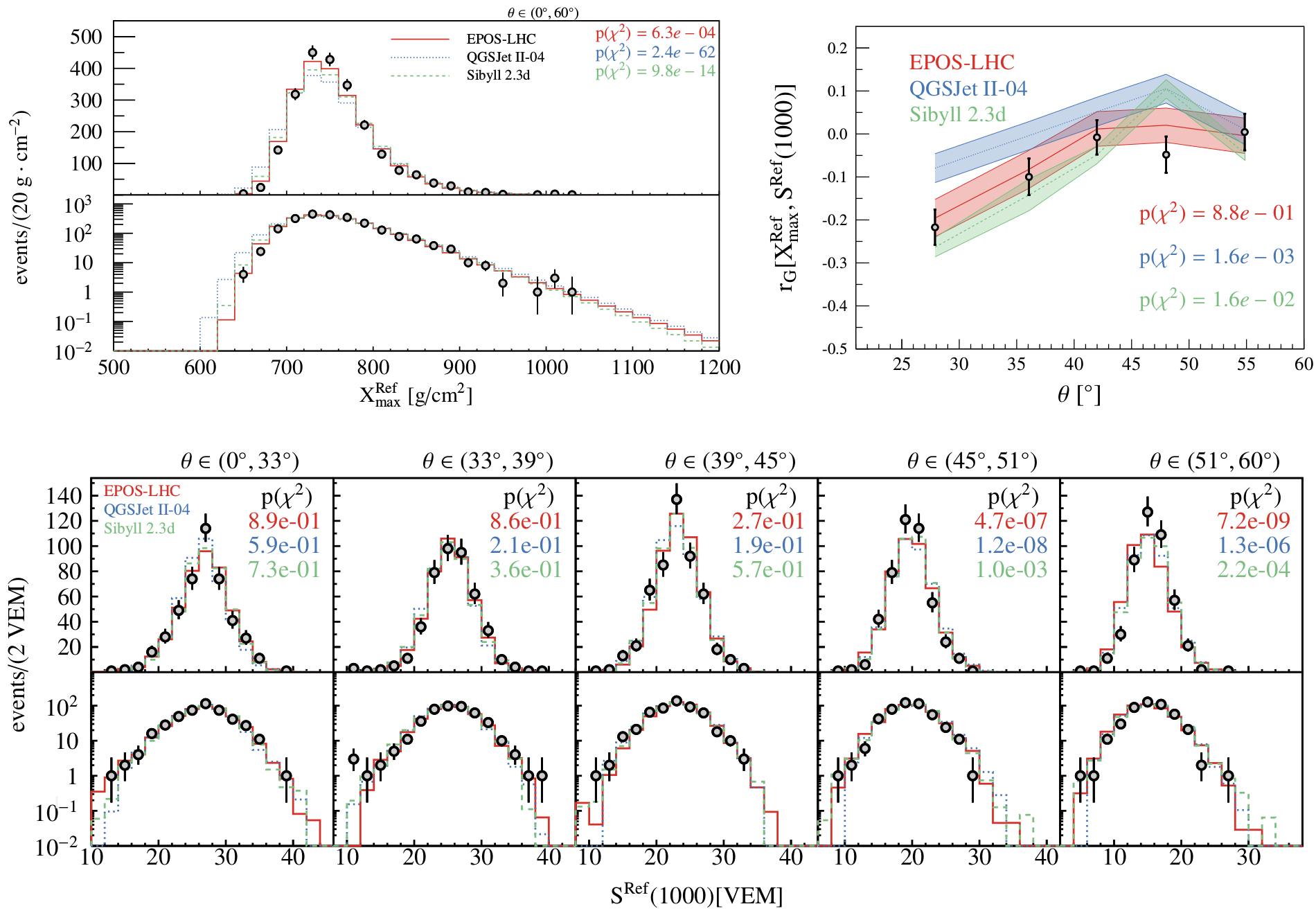}
\caption{Composition fits with a freedom in $R_\text{Had}$, see caption of Fig.~\ref{fig:DataDescription_NoMods}.}
\label{fig:DataDescription_RhadConst}
\end{figure*}

\begin{figure*}
\includegraphics[width=\textwidth]{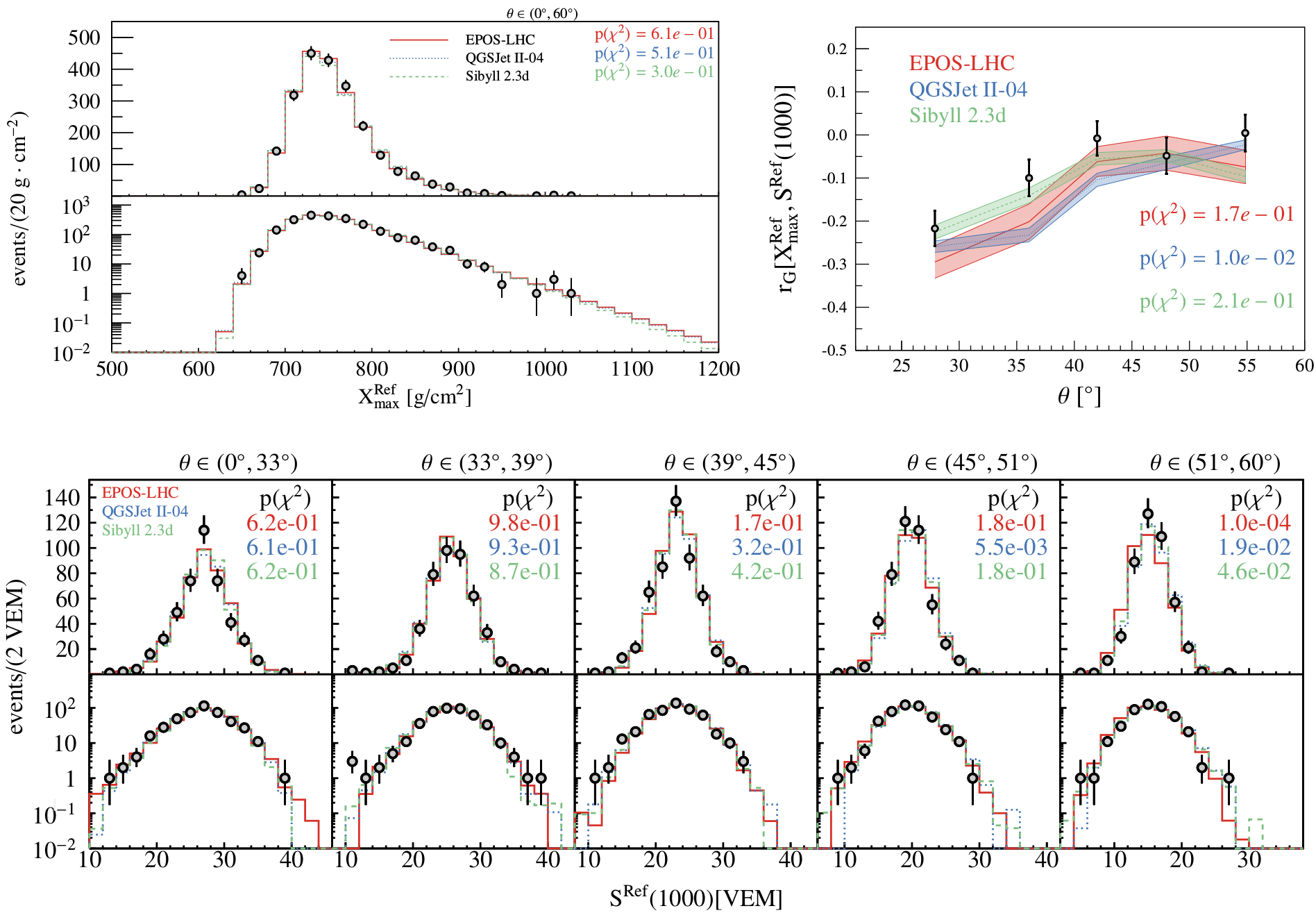}
\caption{Composition fits with a freedom in $R_\text{Had}$ and $\Delta X_\text{max}$, see caption of Fig.~\ref{fig:DataDescription_NoMods}.}
\label{fig:DataDescription_RhadConstAndDeltaXmax}
\end{figure*}

\section{Results}
\label{sec-2}

\subsection{Data Description}
The resulting minimal likelihood-ratio value for different adjustments on the MC predictions is listed in Tab.~\ref{TabLikelihoods} for all three hadronic interaction models.
This value is a measure how well the observed data are described by the adjusted MC templates.
A large improvement in the data description is obtained considering freedom in the predicted hadronic scale and further significant improvement is achieved combining with the freedom in the predicted \Xmax scale as well.
This improvement of data description is further demonstrated in Figs.~\ref{fig:DataDescription_NoMods}, \ref{fig:DataDescription_RhadConst} and \ref{fig:DataDescription_RhadConstAndDeltaXmax} for no adjustments to MC predictions, adjusted hadronic scale and adjusted hadronic and \Xmax scales, respectively.
These figures include projected distributions of the ground signal and \Xmax, together with their zenith-dependent correlation $r_\text{G}$ \citep{rGcoeff}.
The improvement, especially in case of comparison Figs.~\ref{fig:DataDescription_RhadConst} and \ref{fig:DataDescription_RhadConstAndDeltaXmax}, is also quantified by p-values of a $\chi^2$ test.
For all three models, the overall p-value of 2D data description using freedom in both hadronic and \Xmax scales is at a level of several percent based on MC-MC tests.

\subsection{Fitted Parameters}
\begin{figure*}
\subfloat{\includegraphics[width=0.49\textwidth]{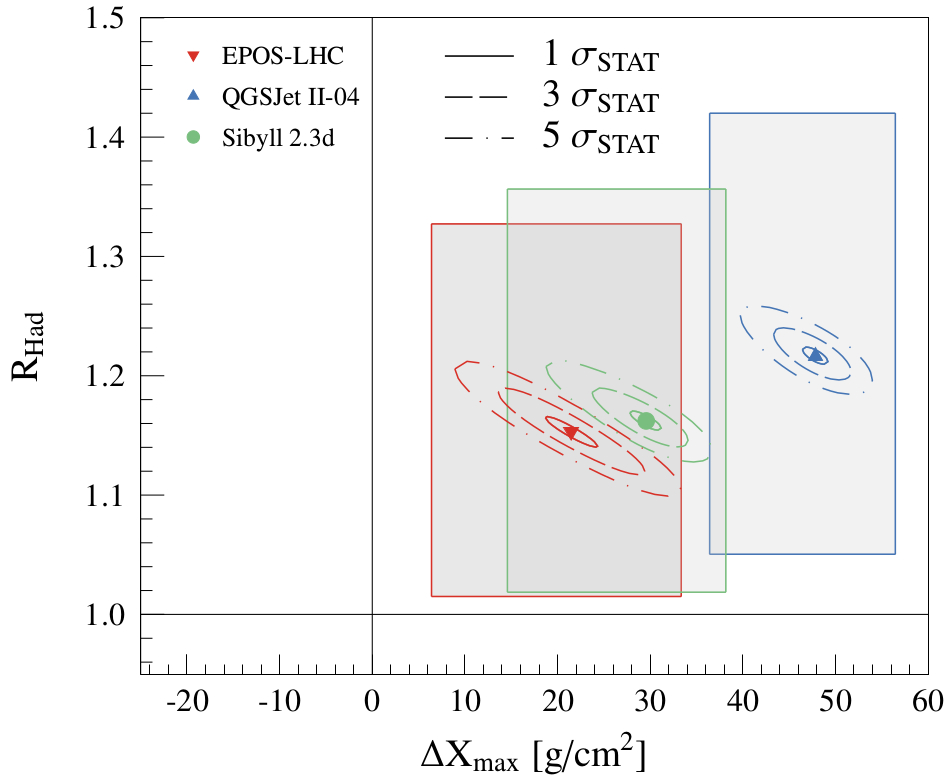}}
\subfloat{\includegraphics[width=0.5\textwidth]{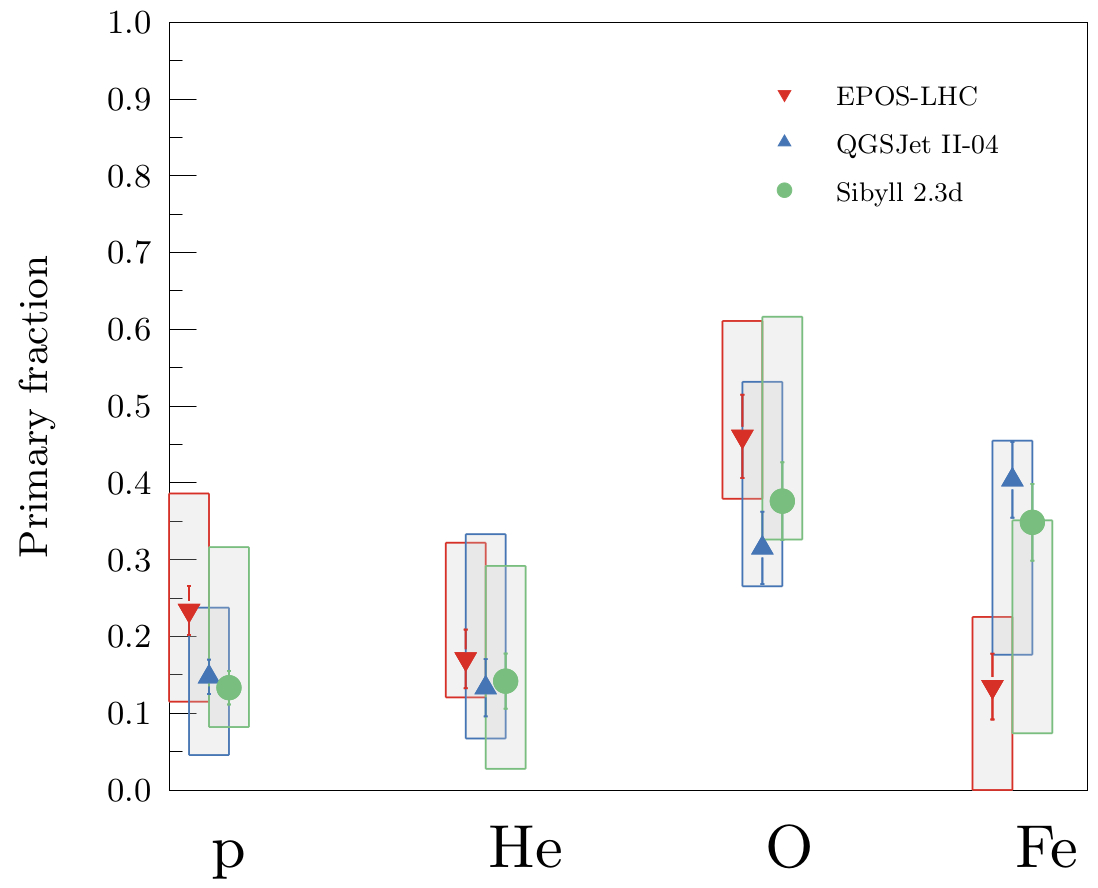}}
\caption{Best-fit values of the fitted parameters $R_\text{Had}$ and $\Delta X_\text{max}$ (left) and primary fractions (right) for all three models of hadronic interactions. The systematic uncertainties are depicted by gray bands.}
\label{fig:McAdjustments}
\end{figure*}

The most likely values of adjustments to the MC templates that describe the Pierre Auger Observatory data are shown on the left panel of Fig.~\ref{fig:McAdjustments}.
The measured data are best described when the MC predictions on \Xmax are on average deeper by $\approx$20, 50 and 30 g/cm$^{2}$ for \eposlhc, \qgsii and \sib{2.3d}, respectively.
In case of the predicted hadronic signal, an increase by 15-25\% is simultaneously needed for all three hadronic interaction models.
The adjusted $X_{\rm max}$ scales have a consequence of heavier mass compositions preferred by the fits to describe the measured data, see the right panel of Fig.~\ref{fig:McAdjustments}, than in case of the mass-composition fits to the measured $X_{\rm max}$ distributions with the unmodified predictions on \Xmax by the hadronic interaction models as in \cite{XmaxFits2014,XmaxIcrc2017} in the corresponding energy range.
As our method removes the main differences in the predictions of models of hadronic interactions by introducing freedom in hadronic and \Xmax scales, the fitted primary fractions show smaller model dependence of the primary fractions than in case of the $X_{\rm max}$ distribution fits with the unmodified predictions.

\subsection{Systematic Uncertainties}
\begin{figure}
\includegraphics[width=0.5\textwidth]{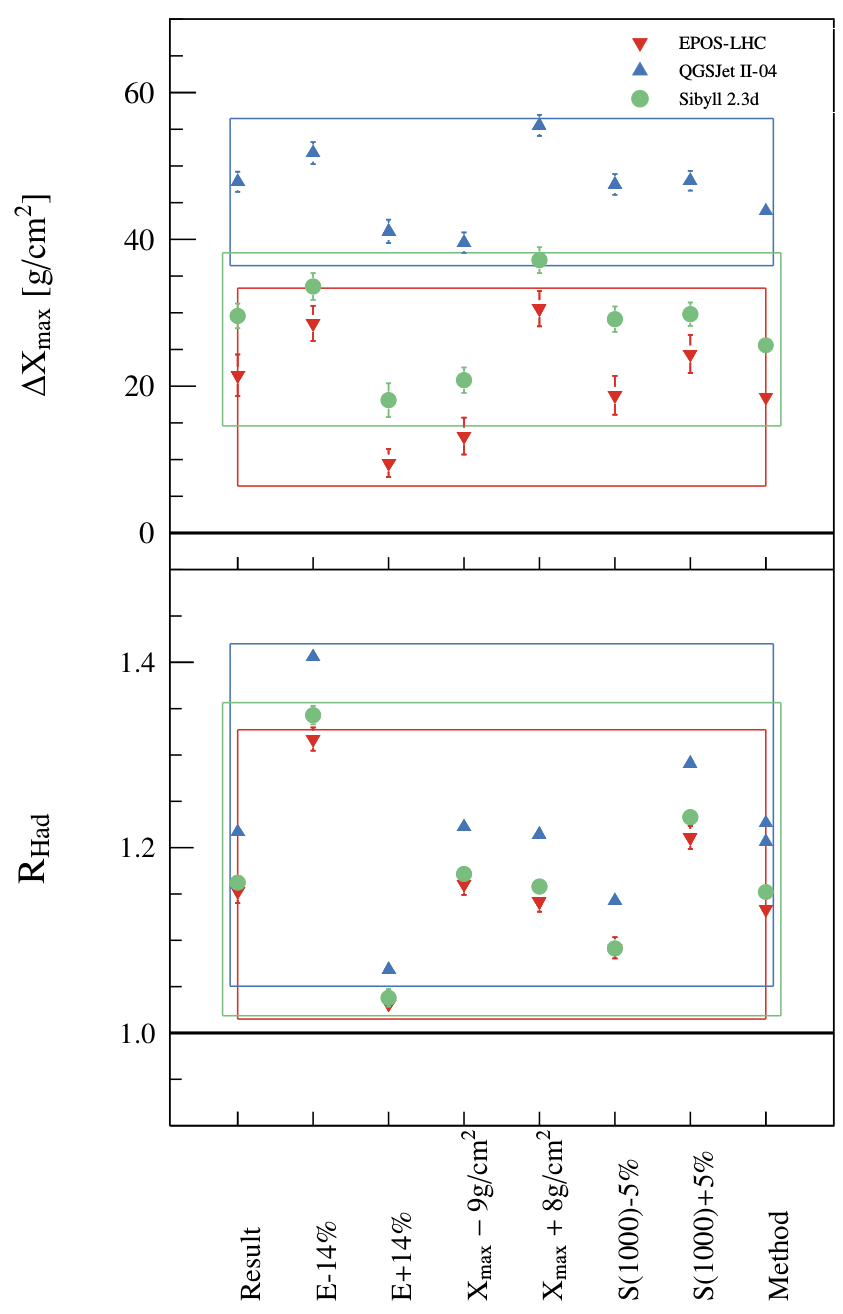}
\caption{Individual systematic contributions (points) to the total systematic uncertainty (band) of $R_\text{Had}$ and $\Delta X_\text{max}$.}
\label{fig:SysContributions}
\end{figure}

There are 4 dominant sources of systematic uncertainties: energy scale ($\pm14\%$), \Xmax measurement ($^{+8}_{-9}$~g/cm$^{2}$), $S(1000)$ measurement ($\pm5\%$) and biases of the method.
The individual contributions of these systematic effects to the MC adjustments are shown in Fig.~\ref{fig:SysContributions} together with their quadratic sums. 

The systematic effects on the energy correction parameters ($B$, $D$, $\beta$) and the long-term effects have negligible contribution compared to the above-mentioned systematic uncertainties as well as the effect of idealization of Auger detectors in the simulations.  

\subsection{Tests of Models}
To quantify the significance of the assumed adjustments to the model predictions, we scan all possible linear combinations of the three experimental systematic uncertainties (energy scale, $X_{\rm max}$ and $S(1000)$ measurements).
The change of attenuation of total ground signal is correlated with the systematic change of energy scale due to different energy dependence of EM ($S_\text{EM}\varpropto E_\text{FD}$) and hadronic ($S_\text{Had}\varpropto E_\text{FD}^{\beta}$) parts of the signal.
The parameter $\beta=0.92$, expressing the growth of hadronic component with energy, is chosen in accordance with \cite{KampertUnger2012}.
Therefore there is a degeneration between the systematic change of energy scale and change of attenuation of the hadronic signal (EM signal is assumed universal in this method, see the right panel of Fig.~\ref{ModelDifferences}).
The correct way to estimate the significance of a need for the assumed MC adjustments is through the more general approach as in \citep{MethodAugerDataICRC21} considering the zenith dependence of $R_\text{Had}$.
Even for the most favorable combination of experimental systematic uncertainties for each model of hadronic interaction, the statistical significance of a need for these MC adjustments is higher than 5$\sigma$.

\section{Conclusions}
\begin{figure*}
\subfloat{\includegraphics[width=0.485\textwidth]{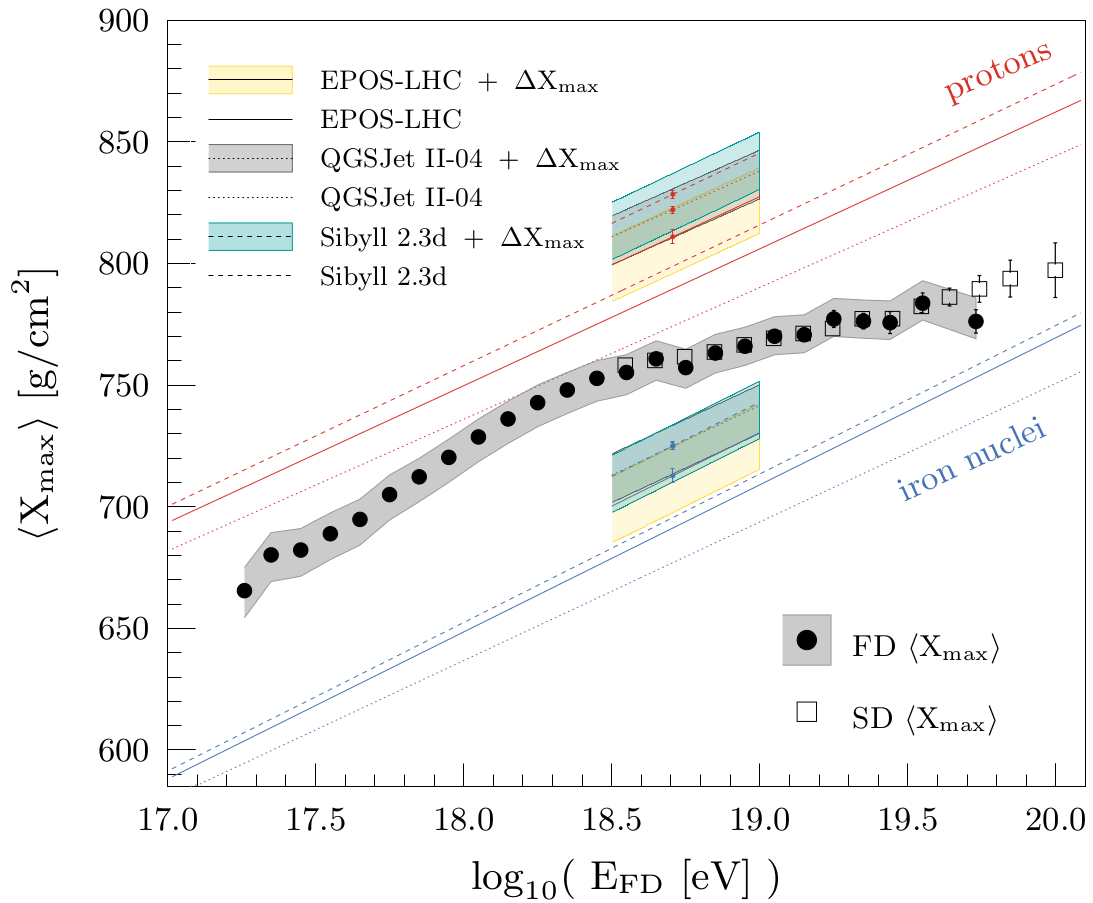}}
\hspace{0.2cm}
\subfloat{\includegraphics[width=0.5\textwidth]{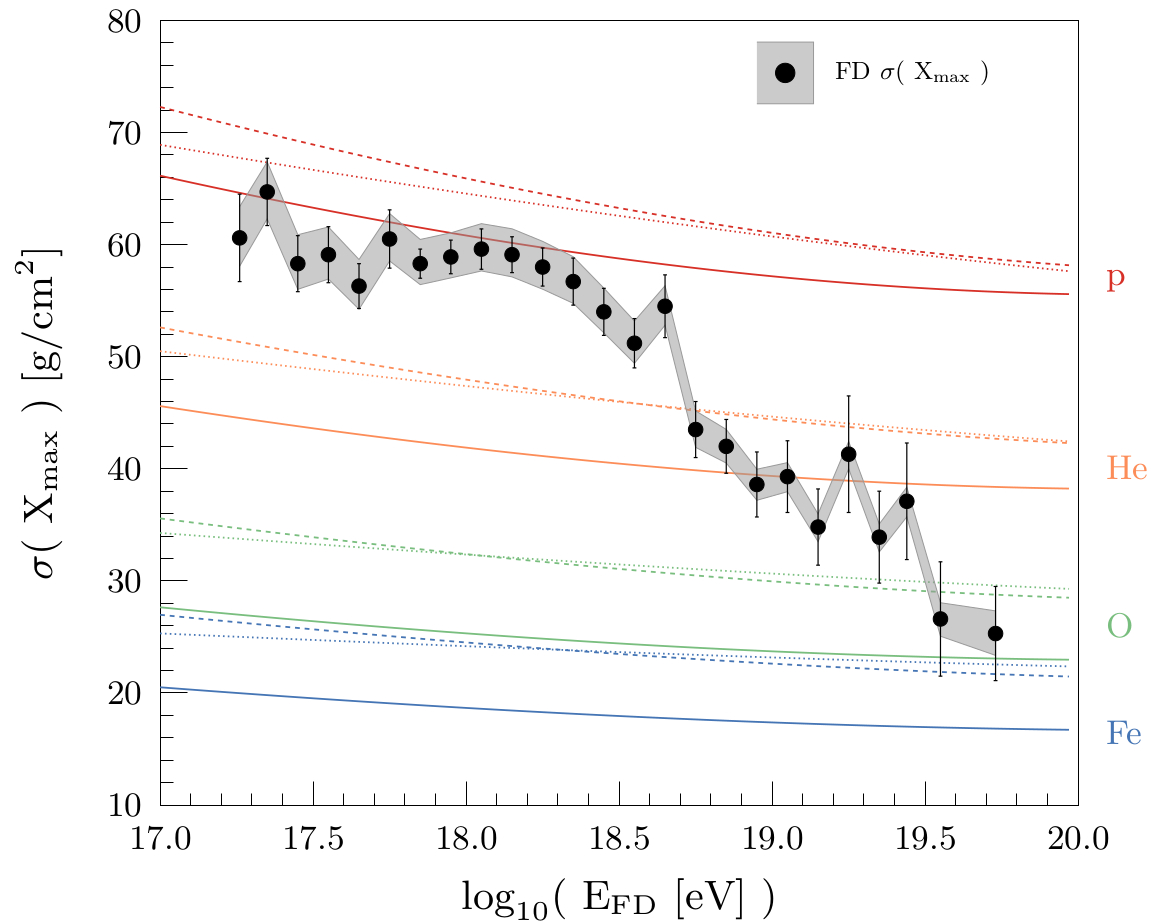}}
\caption{The energy evolution of the mean $X_{\rm max}$ (left) and the standard deviation of $X_{\rm max}$ (right) measured by the Pierre Auger Observatory (in black) using FD \cite{XmaxICRC19} and SD \cite{DeltaICRC19}. The adjusted \Xmax scales of MC predictions (left panel) obtained using the fit results in the energy range \erange{18.5}{19.0} are shown in blue and red for protons and iron nuclei, respectively, with the bands corresponding to the systematic uncertainties. The original predictions by hadronic interaction models are depicted by lines without bands.}
\label{fig:XmaxMoments}
\end{figure*}

The combined measurements of cosmic rays with energies between \lgE{18.5} and \lgE{19.0} using the Surface Detector and the Fluorescence Detector of the Pierre Auger Observatory provide statistics large enough to apply a novel method allowing a complex testing of the air-shower observables predicted by the hadronic interaction models and simultaneously fit the fractions of primary particles.
The results of the method are in tension with the predictions (i.e. $X_{\rm max}$ and $S_\text{Had}$ scales) of hadronic interaction models \eposlhc, \sib{2.3d} and \qgsii by more than 5$\sigma$ with much higher significance in case of the \qgsii.  
The best description of the data is obtained when the $X_{\rm max}$ scale of the MC predictions is deeper by about 20 g/cm$^{2}$, 30 g/cm$^{2}$ and 50 g/cm$^{2}$ for \eposlhc, \sib{2.3d} and \qgsii, respectively.
The adjusted \Xmax scales are shown on the left panel of Fig.~\ref{fig:XmaxMoments} together with measured data.  
At the same time, the differences between the hadronic interaction models in the fitted mass composition fractions are decreased and the mass composition is found heavier than using the unmodified predictions of the three hadronic interaction models.  
For such heavier mass compositions, the "muon problem" of the hadronic interaction models is alleviated with respect to the previous studies, observing $\approx$15-25\% deficit of the hadronic component of simulated ground signal for all three studied models.

\section*{Acknowledgements}
This work was supported by the Ministry of Education, Youth and Sports of the Czech Republic – Grant No. LTT18004, LM2018102 and CZ.02.1.01/0.0/0.0/16\_013/0001402.

\newpage
\bibliography{bibtex}

\begin{thebibliography}{17}

\bibitem{PACosmicObservatory}
{A. Aab et al.} (Pierre Auger Collaboration), {Nucl. Instrum. Methods Phys.
  Res. A} \textbf{798}, 172  (2015)

\bibitem{SDEnergySpectrum2020}
A.~Aab et~al. (Pierre Auger Collaboration), Phys. Rev. D \textbf{102}, 062005
  (2020)

\bibitem{ShowerUniversality}
M.~Ave, R.~Engel, M.~Roth, A.~Schulz, Astropart. Phys. \textbf{87}, 23  (2017)

\bibitem{ShowerUniversality2}
M.~Ave, M.~Roth, A.~Schulz, Astropart. Phys. \textbf{88}, 46 (2017)

\bibitem{Auger-LongXmaxPaper}
A.~{Aab} et~al. (Pierre Auger Collaboration), Phys. Rev. D \textbf{90}, 122005
  (2014)

\bibitem{EposLHC}
T.~Pierog, I.~Karpenko, J.M. Katzy, E.~Yatsenko, K.~Werner, Phys. Rev. C
  \textbf{92}, 034906 (2015)

\bibitem{Qgsjet}
S.~Ostapchenko, Phys. Rev. D \textbf{83}, 014018 (2011)

\bibitem{Sibyll}
F.~Riehn, R.~Engel, A.~Fedynitch, T.K. Gaisser, T.~Stanev, Phys. Rev. D
  \textbf{102}, 063002 (2020)

\bibitem{XmaxICRC19}
A.~Yushkov (Pierre Auger Collaboration) (2019), \texttt{PoS(ICRC19)482}

\bibitem{TestingHadronicInteractions}
A.~Aab et~al. (Pierre Auger Collaboration), Phys. Rev. Lett. \textbf{117},
  192001 (2016)

\bibitem{AmigaMuons}
A.~Aab et~al. (Pierre Auger Collaboration), Eur. Phys. J. C \textbf{210}, 751
  (2020)

\bibitem{MethodAugerDataICRC21}
J.~V\'icha (Pierre Auger Collaboration) (2021), \texttt{PoS(ICRC2021)310}

\bibitem{rGcoeff}
R.A. Gideon, R.A. Hollister, Journal of the American Statistical Association
  \textbf{82}, 656 (1987)

\bibitem{XmaxFits2014}
A.~Aab et~al. (Pierre Auger Collaboration), Phys. Rev. D \textbf{90}, 122006
  (2014)

\bibitem{XmaxIcrc2017}
{J. Bellido} (Pierre Auger Collaboration) (2018), \texttt{PoS(ICRC19)482}

\bibitem{KampertUnger2012}
K.H. Kampert, M.~Unger, Astropart. Phys. \textbf{35}, 660 (2012)

\bibitem{DeltaICRC19}
C.J.T. Peixoto (Pierre Auger Collaboration) (2019), \texttt{PoS(ICRC19)440}

\end{thebibliography}
%
%
%
%

\end{document}